\documentclass[aps,prl,showpacs,twocolumn,amsmath,amssymb]{revtex4}

\usepackage{epsfig}

\begin{document}

\newcommand{\eb}{\epsilon_b}
\newcommand{\ec}{\epsilon_c}
\newcommand{\Ha}{{\mathbf H}}
\newcommand{\M}{{\mathbf M}}
\newcommand{\B}{{\mathbf B}}
\newcommand{\G}{{\mathbf G}}
\newcommand{\ve}{{\mathbf v}}
\newcommand{\La}{{\mathbf L}}
\newcommand{\om}{\omega}
\newcommand{\Om}{\Omega}
\newcommand{\pa}{\partial}
\newcommand{\na}{\nabla}
\newcommand{\taupar}{\tau_{\shortparallel}}
\newcommand{\tauper}{\tau_{\perp}}

\title
{Shape transformations in rotating ferrofluid drops}
\author
{K. I. Morozov$^1$, A. Engel$^2$, A. V. Lebedev$^1$}
\noindent
\affiliation{$^1$ Institute of Mechanics of Continuous Media, Urals Branch of
  the Russian Academy of Sciences, 614013 Perm, Russia\\
 $^2$ Institut f\"ur Theoretische Physik, Otto-von-Guericke Universit\"at,
         PSF 4120, 39016 Magdeburg, Germany}
\begin{abstract}
Floating drops of magnetic fluid can be brought into rotation by applying a
rotating magnetic field. We 
report theoretical and experimental results on the transition from a spheroid
equilibrium shape to non-axissymmetrical three-axes ellipsoids at certain
values of the external field strength. The transitions are continuous for
small values of the magnetic susceptibility and show hysteresis for larger
ones. In the non-axissymmetric shape the rotational motion of the drop
consists of a vortical flow inside the drop combined with a slow rotation of 
the shape. Nonlinear magnetization laws are crucial to obtain quantitative
agreement between theory and experiment.
\end{abstract}

\pacs{47.20.Hw, 47.55.Dz, 75.50.Mm}

\maketitle


The equilibrium shapes of rotating fluid bodies are of importance in various
fields of physics as, e.g., astrophysics \cite{Chandra}, nuclear fission
\cite{BoWh}, plasma \cite{DuON}, and biological physics
\cite{Seifert}. The famous controversy 
between Newton and Cassini on whether the Earth has the form of an oblate or
prolate rotational ellipsoid started a series of ingenious investigations of
the equilibrium shapes of heavenly bodies including work by Maupertius,
MacLaurin, Jacobi, Riemann, Poincar\'e and others. Among the suprising results
discovered are the possibility of three-axes ellipsoids as shown by Jacobi and 
the emergence of pear-shaped configurations found by Poincar\'e. In rotating
non-neutral plasmas, laser cooled in a Penning trap, a novel equilibrium state
with non-axissymmetric surface has recently been observed 
\cite{HBMI}. Tank-treading elliptical membranes in a shear flow have been 
used as a theoretical model for the motion of human red blood cells
\cite{KeSk}. 

Experimental investigations of the stationary shapes of rotating bodies in the
laboratory usually start with a static drop of fluid floating in another,
immiscible fluid of the same density. The rotational motion of the drop is
then set up by, e.g., using a rotating shaft \cite{OhTr} or applying an
acoustic torque \cite{WTCE}. An elegant way to spin up drops made from
polarizable fluids is to use rotating electric \cite{HBMI} or magnetic fields
\cite{BCP,BLP}. 

In the present letter we report theoretical and experimental investigations of
rotating ferrofluid drops and in particular perform the first quantitative
study of a transition from a spheroidal to a non-axissymmetric equilibrium
shape in this system. Moreover, providing an approximate solution of the
hydrodynamic flow problem inside and outside the drop we are able to analyze
the rotational motion of the drop shape and to separate it from the internal
hydrodynamic flow.  

Ferrofluids are suspensions of ferromagnetic nano-particles in suitable
carrier liquids combining the hydrodynamic behaviour of a Newtonian fluid with
the magnetic properties of a super-paramagnet \cite{Ro}. A rotating external
field induces a rotational motion of the nano-particles which due to
their viscous coupling to the surrounding liquid transfer the angular momentum
to the whole drop. The system has been studied previously by Bacri et~al. 
using microdrops with a typical radius of 10 $\mu m$, very small surface 
tension and a viscosity much larger than that of the outer fluid 
\cite{BCP,JEBCP}. Although an instability of the axissymmetric shape was found
experimentally and theoretically \cite{Cebers} the emerging
non-axissymmetric configurations could not be studied quantitatively since for
the given parameter values very irregular shapes arise. In our 
experiments we used much larger drops of radius $R=2.75$mm of a kerosine-based
ferrofluid with magnetic susceptibility $\mu_r=15.3$ and dynamic viscosity 
$\eta_1=0.019$ Pas immersed in 3-brome-1,2-propandiol with dynamic viscosity 
$\eta_2=0.058$ Pas. The interface tension is $\sigma=2.8\cdot 10^{-3}$ N/m, 
the frequency of the magnetic field 560 Hz. For these
parameter combinations we find well-defined transitions to three-axes
ellipsoids which can be analyzed in detail. 


From the parameter values given above we find for the typical time of shape
relaxations of the drop $\tau_s\cong 0.1$s. We are thus considering the case
of a fast rotating field, $\om\tau_s\gg 1$, in which the drop assumes an
oblate shape with the short axis perpendicular to the field plane. Its form is
very near to an ellipsoid and we will use this approximation throughout our
theoretical analysis. The shape will be specified by the ratios
$\eb=a/b$ and $\ec=a/c$ between the seminaxes of the ellipsoid where as usual
$a\ge b\ge c$ is assumed. 

Estimating the orders of magnitude of the energies due to magnetization,
surface tension, viscosity and intertia, respectively, we find that for the
experimentally relevant parameter values the shape is almost completely 
determined by the balance between surface and magnetic energy where the latter
is conveniently averaged over one period of the field rotation. The
calculation of the magnetic energy is simplified by the fact that an ellipsoid
in a homogenous external field builds up a {\it homogenous}
magnetization. Using a coordinate system in which the 
external field is of the form $\G=(G \cos(\om t),G\sin(\om t),0)$ this 
magnetization is determined by the relations 
\begin{equation}\label{GHM}
  G_x=H_x+n_x M_x \quad,\quad G_y=H_y+n_y M_y
\end{equation}
between the components of the external field $\G$, the internal field $\Ha$
and the magnetization $\M$ \cite{LL}. Here $n_x$ and $n_y$ denote the
demagnetization factors along the $x$ and $y$ axis, respectively, which are
known functions of $\eb$ and $\ec$ \cite{LL}. 

\begin{figure}[t]
\includegraphics[width=8cm]{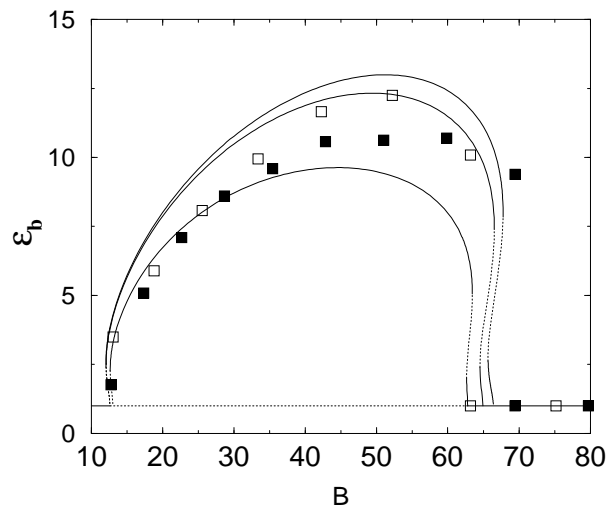} 
   \caption{\label{fig1} Ratio $\eb=a/b$ between the two largest semiaxes
     of a rotating ferrofluid drop as a function of the magnetic Bond number
     $B$. Squares are experimental values with filled symbols corresponding to
     increasing, empty symbols to decreasing field strength, respectively. The
     curves show from top to bottom the results for a linear 
     magnetization law, $\M=\chi\Ha$, for the Langevin $M(H)$ and for 
     the dynamic curve $M(H)$ as determined 
     from an independent experiment. Full lines correspond to stable
     configurations, dotted lines to unstable ones. There are no fit
     parameters.} 
\end{figure}
\begin{figure}[t]
\includegraphics[width=8cm]{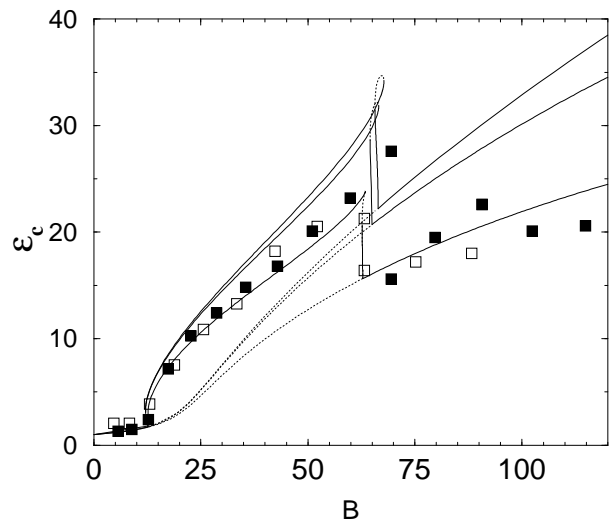} 
   \caption{\label{fig2} Same as fig.\ref{fig1} for the ratio $\ec=a/c$
     between the largest and the smallest semiaxes of the rotating drop.}
\end{figure}

The calculation of the magnetic energy is most easily accomplished
assuming a linear magnetization law, $\M=\chi \Ha$, with the susceptibility
$\chi=\mu_r-1$. We then find for the magnetic energy 
$E_m=-\mu_0 V\M\cdot\G/2$ using (\ref{GHM}) 
\begin{equation}\label{magenlin}
E_m(t)=-\frac{\mu_0 V}{2} \chi G^2 \left(\frac{\cos^2(\om t)}{1+\chi n_x}+
       \frac{\sin^2(\om t)}{1+\chi n_y}\right)
\end{equation}
with $V$ denoting the volume of the drop. Averaging over one period of the
rotation \cite{rem1}, using the well-known expression for the surface of a
three-axes ellipsoid, and observing volume conservation for the drop we obtain
the following result for the sum $E$ of magnetic and surface energies
\begin{multline}\label{enlin}
\frac{E}{2\pi\sigma R^2}=
-\frac{\chi}{6}\; B \left(\frac{1}{1+\chi n_x}+\frac{1}{1+\chi n_y}\right) + \\
  \eb^{2/3} \ec^{-4/3}\left[1 + 
  \frac{\ec}{\eb\sqrt{\ec^2-1}}\left(F(m,\kappa)+(\ec^2-1)E(m,\kappa)\right)
  \right]\nonumber
\end{multline}
Here $F$ and $E$ are elliptic integrals of first and second kind,
respectively, 
\cite{AbSt}, $m=\sqrt{\ec^2-1}/\ec$, $\kappa=\sqrt{(\ec^2-\eb^2)/(\ec^2-1)}$,
and $B=\mu_0 G^2 R/\sigma$ is the magnetic Bond number measuring the strength
of the external field.  
Minimizing this expression numerically in the geometry parameters $\eb$ and
$\ec$ the dependence of the shape of the drop can be determined for varying
external field strength $G$. Results for the parameter values given above are
shown in figs.\ref{fig1} and \ref{fig2} together with our experimental
findings. 

The main result is a transition from a spheroid characterized by $\eb=1$ to
a pronounced non-axissymmetric form of a {\it three-axes ellipsoid} for
intermediate values of the magnetic field strength. This transition takes
place only if the magnetic permeability is 
large enough, $\mu_r\gtrsim 5.1$. It occurs via supercritical bifurcations up
to $\mu_r\cong 11.6$ and through subcritical transitions with hysteresis for
still higher values of $\mu_r$ including our experimental value. 

From the figures it is seen that the linear theory 
is in qualitative agreement with the experiment. It also yields 
quantitatively good results for the values of the Bond number at
which the transitions to three-axes ellipsoids occur. 
The ratios between the semiaxes, however, are overestimated, with the
discrepancy increasing with the field strength. In fact the magnetic field
corresponding to a Bond number of $B\cong 80$ is of the order of the
saturation magnetization $M_{\infty}\cong 80$kA/m and hence deviations
from the linear law $\M=\chi\Ha$ become important. 

In order to account for these deviations from a linearity we have extended the
theory to general magnetization curves $M(H)$. The field inside the ellipsoid
is still homogeneous and using (\ref{GHM}) we get 
\begin{equation}\label{eqH}
  \frac{1}{G^2}=\frac{\cos^2(\om t)}{(H+n_x M(H))^2}+
                \frac{\sin^2(\om t)}{(H+n_y M(H))^2}
\end{equation}
from which for given $M(H)$ the time dependence of the internal field $H$ can
be determined numerically. The magnetic energy can now be written in the form 
\begin{align}\label{magennonlin}
  E_m(t)=&-\frac{\mu_0 V}{2}
   \left[G^2 M\left(\frac{\cos^2(\om t)}{H+n_x M}+
          \frac{\sin^2(\om t)}{H+n_y M}\right)\right.\nonumber\\
         & \left.- MH +2 \int_0^H dH' M(H')\right].
\end{align}
For a linear magnetization law the last two terms cancel and we find back
(\ref{magenlin}). Expression (\ref{magennonlin}) is numerically averaged over
one period of the field rotation and the total energy is minimized with
respect to $\eb$ and $\ec$.  

A standard choice for the magnetization curve of a ferrofluid is the Langevin
law $M=M_{\infty}(\coth(\xi)-1/\xi)$ with $\xi=3\chi H/M_{\infty}$. The
corresponding results are included in figs.\ref{fig1} and \ref{fig2}. 
It is clearly seen that although the values for $\eb$ and $\ec$ are reduced
in comparison with the linear theory the corrections are rather small. 
The physical reason for the remaining discrepancy lies in relaxation effects
of the magnetization which for the comparetively high frequency of the
external field start to play a role. We have therefore determined the
magnetization curve $M(H)$ of our ferrofluid for an 
oscillating magnetic field from an independent experiment. The lowest curves
in figs.\ref{fig1} and \ref{fig2} result from the use of a spline
interpolation of these data in our numerical code. Whereas the experimental
results for $\eb$ are now underestimated for large values of $B$ the
correspondence between theory and experiment for $\ec$ is rather satisfactory.


We now turn to the analysis of the drop motion. In the experiments we observe 
a slow rotation of the non-axissymmetric drop shape with angular velocity
$\Om\cong 0.1$Hz in the direction of the field rotation. In order to discuss
this motion theoretically 
we have to solve the hydrodynamic flow problem inside and outside the drop
taking into account the appropriate boundary conditions for normal and
tangential stresses at the surface. In its general form this is a formidable
free boundary value problem. We hence employ some reasonable approximations 
consistent with our assumption of an elliptical drop shape. For the
experimentally relevant sizes and viscosities the inertial terms in the
hydrodynamic equations are small and we may use the Stokes approximation for
the determination of the flow fields. It is convenient to use the coordinate
system in which the shape of the drop is at rest. In this frame we 
assume that the motion inside the drop is of uniform vorticity $\zeta$ and
hence use for the internal and external flow fields the ansatzes
\begin{equation}\label{vansatz}
  \ve^{\text{in}}=(-\zeta \frac{y a}{b},\;\zeta \frac{x b}{a},\;0),\quad
    \ve^{\text{ex}}=\ve^{\text{in}}+(u_J,v_J,w_J).
\end{equation}
From the continuity of the velocities at the surface of the drop we infer that
$(u_J,v_J,w_J)$ describes the flow field outside a {\it rigid} elliptical
particle at rest in a viscous fluid with asymptotic velocity 
$\lim_{r\to\infty} (u_J,v_J,w_J)=((\Om+\zeta a/b) y,-(\Om+\zeta b/a) x,0)$, 
a problem solved by Jeffery many years ago \cite{Jeffery}.

To complete the solution in the discussed approximation we need two equations
to fix the so far undetermined parameters $\Om$ and $\zeta$. As first equation
we use the balance between the viscous 
torque experienced by the drop in its rotational motion and the magnetic
torque, again averaged over one period of the driving. The second one results
from the stationarity of the energy balance stating that the average work done
by the external field per unit time must be equal to the energy dissipated per
unit time in the viscous flows. This latter reqirement can be rewritten as a
continuity condition for the tangential stress {\it averaged} over the 
surface of the drop \cite{Morozov,rem2}. 

The expression for the viscous torque builds on Jeffery's solution and can be
obtained in the same way as for a slowly rotating field, $\om \tau_s\ll 1$,
\cite{LeMo}. For the determination of the magnetic torque 
$\La= \mu_0 V\, (\M\times\G)$ it is crucial to take into account the finite
relaxation time of the magnetization since a non-zero averaged torque results
from the phase lag between  
$\M$ and $\Ha$. The explicit calculation is again feasible analytically for a
linear magnetization curve $\M=\chi\Ha$ where the susceptibility is now
complex, $\chi=\chi_1-i\chi_2$. We will assume $\chi_2\ll\chi_1$ consistent
with the experiment where $\chi_1\cong 14.3$ and $\chi_2\cong 2.1$. Using
(\ref{GHM}) and $\om-\Om\cong \om$ we get for the $z$-component 
of the averaged magnetic torque  
\begin{equation}\label{magtorquelin}
  \overline{L_z}=\frac{\mu_0 V G^2}{2} \chi_2 
    \left(\frac{1}{(1+\chi n_x)^2}+\frac{1}{(1+\chi n_y)^2}\right).
\end{equation}
In the integrated balance of tangential stresses the viscous contributions
can be obtained from Jeffery's solution \cite{Jeffery,Morozov}. The tangential
part of the Maxwell stress tensor of the magnetic field is automatically
continuous at the interface due to the boundary conditions obeyed by the fields
\cite{LL}. However, in the present non-equilibrium situation there {\it is}
a magnetic contribution due to the additional term $(M_i H_k-M_k H_i)/2$ in
the stress tensor introduced by Shliomis \cite{Sh}. Taking into account the
various contributions two equations for $\Om$ and $\zeta$ can be derived. The
final expressions are rather long and will be published elsewhere \cite{ELM}.

\begin{figure}[tb]
\includegraphics[width=8cm]{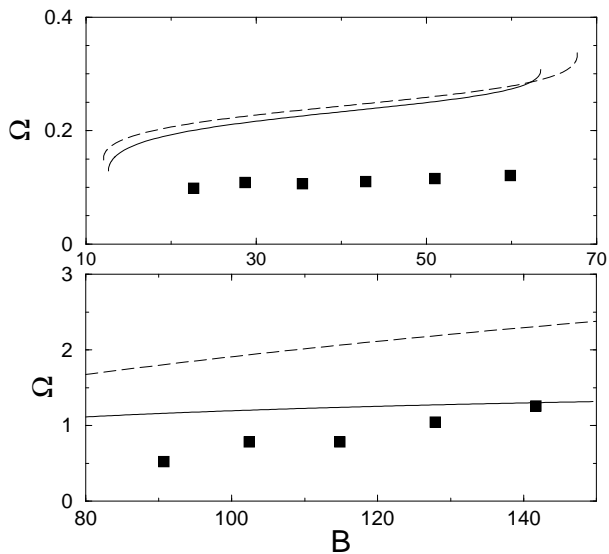} 
   \caption{\label{fig3} Rotation frequency of the drop as function of the
     magnetic Bond number. Squares are experimental values, lines are
     theoretical results using a linear magnetization law (dashed) and the
     magnetization curve $M(H)$ as determined in an independent experiment
     (full), respectively. The upper part describes the rotation of the
     three-axes ellipsoid, the lower one the hard body rotation of a disk.}
\end{figure}

Results for the rotation frequency $\Om$ of the drop are shown in
fig.\ref{fig3} together with the corresponding experimental findings. The
linear theory overestimates the magnetic torque and therefore also the
frequency of rotation. To investigate the effects of deviations from the linear
magnetization curve we have calculated the magnetic torque using the numerical
solution of the magnetization relaxation equation \cite{MRSh}
\begin{equation}
  \frac{d \M}{d t}=-\frac{1}{\taupar H^2}(\Ha\cdot(\M-\M_0))\Ha
                  -\frac{1}{\tauper H^2}\Ha\times(\M\times\Ha).\nonumber
\end{equation}
The static magnetization $\M_0(H)$ and the field dependence of the relaxation
times $\taupar$ and $\tauper$ are determined from independent experiments, the
geometry of the drop is taken from figs.\ref{fig1} and \ref{fig2}. Although 
the magnetic torque is substantially smaller than in the linear case the
results for the rotation period of the non-axissymmetric drop are almost the
same. This is due to the reduced viscous torque resulting from smaller values
of $\eb$ and $\ec$ in the nonlinear theory. The remaining differences with the
experiment may be due to the polydispersity of the fluid requiring a whole
spectrum of relaxation times and a small ellipticity of the external field. 
These questions will be dealt with in detail elsewhere \cite{ELM}.

For the axissymmetric state, $\eb=1$, shown in the lower part of
fig.\ref{fig3} the situation is somewhat simpler. In this case the drop forms
a flat disk rotating like a hard body. For large values of $B$ this disk
developes peaks around its perimenter due to the normal field instability
\cite{Ro,BCP,Cebers} making an experimental determination of the rotation
frequency 
easy. As can be seen from fig.\ref{fig3} the linear theory again strongly
overestimates the rotation frequency whereas the non-linear theory yields
substantially smaller results. For Bond numbers between 80 and 100 the disk
shows few, rather large peaks resulting in an increased viscous torque and
therefore a slower rotation. For larger
values of the Bond number the peaks get smaller and more numerous and the
approximation of the shape by an ellipsoid becomes again more accurate
resulting in a reasonable agreement between theory and experiment.\\

\begin{acknowledgments}
 We would like to thank Marcus Hauser for his help in procuring some of the
 chemicals needed for the experiments. Financial support from the 
{\it Deutsche Forschungsgemeinschaft} under grant FOR 301/2-1 is gratefully
acknowledged. 
\end{acknowledgments}


\begin{thebibliography}{99}
\bibitem{rem1} We assume $n_x$ and $n_y$ as being constant
  anticipating that the rotation of the elliptic shape of the droplet is much
  slower than that of the external magnetic field.
\bibitem{rem2} Note that it is impossible within our approximations for the
  shape and the flow fields to fulfill this continuity at every point of the 
  surface independently.
\end{thebibliography}
\end{document}